\documentclass[fleqn,usenatbib]{mnras}

\usepackage{newtxtext,newtxmath}

\usepackage[T1]{fontenc}

\DeclareRobustCommand{\VAN}[3]{#2}
\let\VANthebibliography\thebibliography
\def\thebibliography{\DeclareRobustCommand{\VAN}[3]{##3}\VANthebibliography}

\usepackage{graphicx}	
\usepackage{amsmath}	
\usepackage{enumitem}

\newcommand{\kappaco}{\kappa_\mathrm{co}}
\newcommand{\tspin}{t_\mathrm{spin}}

\title[Milky Way disc formation in \textsc{artemis}]{Taking the Milky Way for a spin: disc formation in the \textsc{artemis} simulations}

\author[A. M. Dillamore et al.]{
Adam M. Dillamore,$^{1}$\thanks{E-mail: amd206@cam.ac.uk (AMD)}
Vasily Belokurov,$^{1}$
Andrey Kravtsov$^{2,3,4}$
and Andreea S. Font$^{5}$
\\
$^{1}$Institute of Astronomy, University of Cambridge, Madingley Road, Cambridge CB3 0HA, UK\\
$^{2}$Department of Astronomy and Astrophysics, The University of Chicago, Chicago, IL 60637 USA\\
$^{3}$Kavli Institute for Cosmological Physics, The University of Chicago, Chicago, IL 60637 USA\\
$^{4}$Enrico Fermi Institute, The University of Chicago, Chicago, IL 60637\\
$^{5}$Astrophysics Research Institute, Liverpool John Moores University, 146 Brownlow Hill, Liverpool L53RF, UK
}

\date{Accepted XXX. Received YYY; in original form ZZZ}

\pubyear{2023}

\begin{document}
\label{firstpage}
\pagerange{\pageref{firstpage}--\pageref{lastpage}}
\maketitle

\begin{abstract}
We investigate the formation (spin-up) of galactic discs in the \textsc{artemis} simulations of Milky Way-mass galaxies. In almost all galaxies discs spin up at higher [Fe/H] than the Milky Way (MW). Those that contain an analogue of the Gaia Sausage-Enceladus (GSE) spin up at a lower average metallicity than those without. We identify six galaxies with spin-up metallicity similar to that of the MW, which form their discs $\sim 8-11$~Gyr ago. Five of these experience a merger similar to the GSE. The spin-up times correlate with the halo masses at early times: galaxies with early spin-up have larger virial masses at a lookback time $t_L=12$~Gyr. The fraction of stars accreted from outside the host galaxy is smaller in galaxies with earlier spin-ups. Accreted fractions small enough to be comparable to the MW are only found in galaxies with the earliest disc formation and large initial virial masses ($M_\mathrm{200c} \approx2\times10^{11}M_\odot$ at $t_L=12$ Gyr). We find that discs form when the halo's virial mass reaches a threshold of $M_\mathrm{200c}\approx(6\pm3)\times10^{11}M_\odot$, independent of the spin-up time. However, the failure to form a disc in other galaxies appears to be instead related to mergers at early times. We also find that discs form when the central potential is not particularly steep. Our results indicate that the MW assembled its mass and formed its disc earlier than the average galaxy of a similar mass.

\end{abstract}




\section{Introduction}

The Milky Way, like all other spiral galaxies, was not born with a rotating disc, it had to form one. Exactly when and how galaxies acquire stable stellar discs is an open question, currently pursued both with cutting-edge  look-back observations and through theoretical and numerical calculations.  

The ratio of the rotational velocity to the velocity dispersion $v/\sigma$ is measured to decrease significantly with increasing redshift for galaxies of a given stellar mass \citep[see][]{Kassin12,Wisnioski15}. This disc {\it settling} is a mass-dependent process: more massive galaxies attain higher $v/\sigma$ values earlier \citep[][]{Wisnioski19}. Mixed in with these trends is the information on the disc {\it emergence}, i.e. the galaxy's transformation from a state characterised by random motions to a configuration dominated by a coherently rotating stellar disc. Currently, available samples of high-redshift galaxies with morphological and kinematic measurements are still too small to answer this question through population studies \citep[but see recent studies by][]{Ferreira.etal.2022,Ferreira.etal.2023,Nelson23,Jacobs.etal.2023,Robertson.etal.2023}. However, in the Milky Way, archaeological records comprised of accurate kinematic information provided by {\it Gaia} \citep[][]{Gaia} and detailed chemical abundances supplied by surveys such as APOGEE \citep[][]{apogee} and GALAH \citep[][]{Galah} have recently been used to crack the puzzle of the disc emergence.

For example, \citet{belokurov22} show that the stars born in-situ in the Milky Way (separated from the accreted stellar debris using the abundance ratio of aluminium to iron [Al/Fe]) exhibit a characteristic trend of the increasing rotational velocity with metallicity [Fe/H]. At low metallicity, i.e. at [Fe/H]~$<-1.3$, the Milky Way stars possess little to no net spin, however from [Fe/H]$\approx-1.3$ to [Fe/H]~$\approx-0.9$, the median azimuthal velocity $v_{\phi}$ rises from approximately $\approx50$ km/s to $\approx150$ km/s. \citet{belokurov22} associate this rapid {\it spin-up} with the emergence of the old Galactic disc, following the turbulent and chaotic state of the Galaxy imprinted in the ancient in-situ stellar population {\it Aurora}.\footnote{Named after Aurora -- the Latin name of the goddess of dawn Eos in Greek mythology.} 
Whilst more contaminated in their selection of the in-situ population, the follow-up studies of \citet{Conroy22} and \citet{Rix22} report similar trends of the rotational velocity with metallicity. 

\citet{belokurov22} also highlight a discrepancy between the observations and the numerical simulations of galaxy formation. Whilst the spin-up itself is a ubiquitous feature in chemo-kinematic histories of model Milky Way-sized galaxies, the metallicity at which the disc forms in the two simulation suites analysed (Auriga and FIRE) is noticeably higher compared to the observations. Similar results were also produced by \citet{mccluskey23} using the FIRE-2 simulations, although one of these galaxies (named Romeo) does have a similar time of disc formation to the Milky Way. This question is further explored in \citet{semenov23} using a larger sample of Milky Way-sized galaxies from the Illustris TNG50 simulation to study the statistics of the spin-up metallicity. They conclude that in TNG50 only $\approx 10\%$ of Milky Way-sized galaxies form their disc at metallicities similar to that measured by \citet{belokurov22}. They also show that the haloes of these galaxies assemble their mass early. 

One of the most significant events in the Milky Way's history was the merger with \textit{Gaia} Sausage-Enceladus \citep[GSE;][]{belokurov18,helmi18}, a massive satellite of total mass $\sim10^{11}M_\odot$ which was accreted by the Milky Way 8-11 Gyr ago \citep[e.g.][]{belokurov18,Fattahi}. This is observed as a population of stars in the stellar halo with highly eccentric orbits and relatively high metallicity. Milky Way analogues with GSE-like mergers have been studied in various suites of high resolution zoomed-in cosmological simulations, such as Auriga \citep{grand2017,Fattahi,grand2020} and \textsc{artemis} \citep{artemis,dillamore22_artemis,dillamore23_artemis}. These have shown that approximately one third of Milky Way-like galaxies possess a dominant radially anisotropic feature in their stellar haloes resembling the GSE \citep{Fattahi}, and that the associated mergers have transformative effects on their host galaxies \citep{dillamore22_artemis}. These include creation of an \textit{in-situ} stellar halo \citep{belokurov2020,grand2020} and tipping of the disc \citep{dillamore22_artemis,orkney2023}.

Following the above early attempts to gauge the conditions necessary for the disc emergence, in this study we focus on the connection between the disc spin-up in model Milky Way-like galaxies and their mass assembly histories using the \textsc{artemis} simulation suite. This paper is arranged as follows. We briefly describe the \textsc{artemis} simulations in Section~\ref{section:simulations} and our methods in Section~\ref{section:method}. Our results are presented and discussed in Section~\ref{section:results}, and summarised in Section~\ref{section:summary}. Finally, in Appendix~\ref{section:SFR} we show the star formation density of a selection of \textsc{artemis} galaxies across time to illustrate the changes in morphology involved in disc formation.

\section{Simulations}
\label{section:simulations}

\textsc{artemis} \citep{artemis} is a zoomed-in hydrodynamical simulations suite, consisting of 45 haloes with Milky Way-like masses. Details can be found in \citet{artemis} and are outlined below.

The simulations were run using the Gadget-3 code \citep{springel05} in a WMAP $\Lambda$CDM cosmology. While the hydrodynamics code and subgrid physics are shared with the \textsc{eagle} project \citep{schaye15,crain15}, the stellar feedback was recalibrated to better match the stellar mass-halo mass relation \citep{artemis}.

An initial collisionless simulation was run to redshift zero in a base periodic box of side length 25 Mpc $h^{-1}$. A set of 45 galaxies was then selected based purely on their virial masses. These lie in the range $0.8<M_{\rm 200c}/10^{12}\,M_\odot<2.0$  at redshift $z=0$, where $M_{\rm 200c}$ is the mass enclosed within a volume containing a mean density of 200 times the critical density.

The final high-resolution simulations have a dark matter particle mass of $1.17\times10^5\,h^{-1}M_\odot $ and an intial baryon particle mass of $2.23\times10^4\,h^{-1}M_\odot $. The Plummer-equivalent softening length is $125\,h^{-1}\rm pc$. Haloes and subhaloes were identified with the SUBFIND algorithm \citep{Dolag_subfind}.

The size-stellar mass relation of the \textsc{artemis} galaxies closely matches the observations by \citet{shen03} in the SDSS survey \citep{artemis}. However, the typical disc masses of the \textsc{artemis} galaxies are $\approx10^{10}M_\odot$, smaller than the estimate for the Milky Way of $\approx (5\pm 0.5)\times10^{10}M_\odot$ \citep[e.g.,][]{mcmillan17}.

\begin{figure*}
  \centering
  \includegraphics[width=0.95\textwidth]{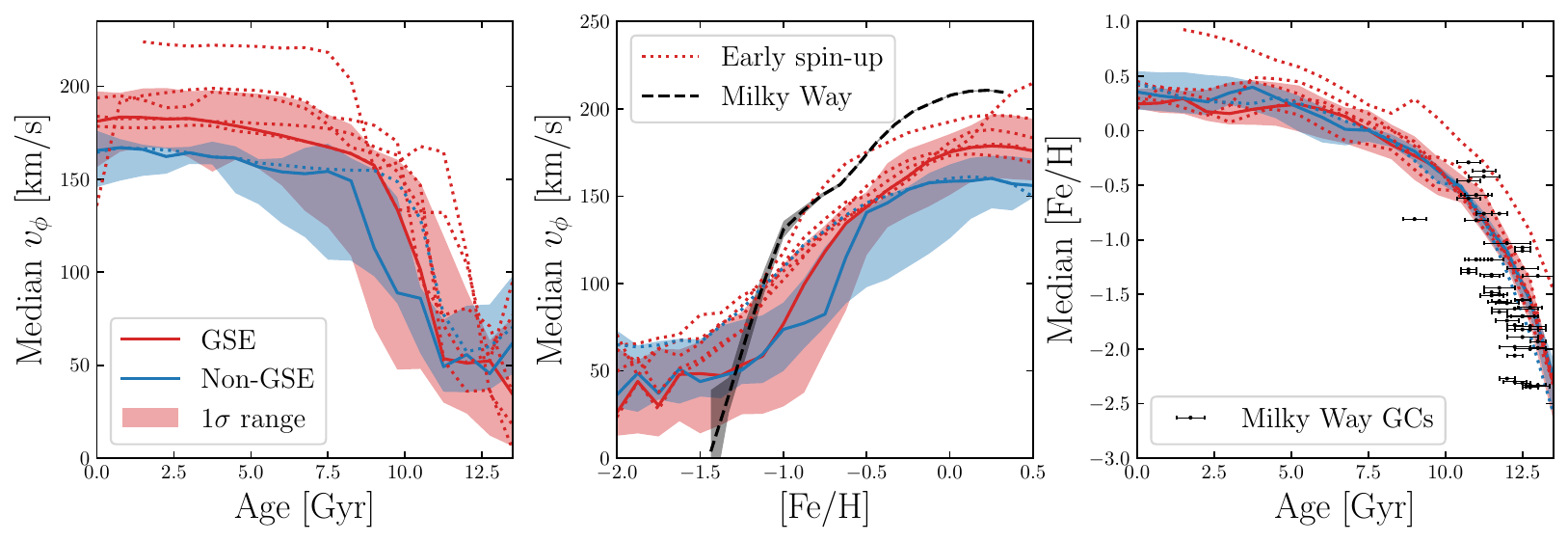}
  \caption{\textbf{Left-hand panel:} median $v_\phi$ as a function of age, with red (blue) indicating GSE (non-GSE) galaxies. The solid lines and bands indicate the medians and 16th-84th percentile ranges of each sample. \textbf{Middle panel:} as above, but as a function of [Fe/H]. Data from the Milky Way is shown with the black dashed line. The dotted lines mark the individual \textsc{artemis} galaxies selected as `early spin-up', those with the highest median $v_\phi$ at [Fe/H]~$=-1$. These have similar spin-up profiles to the Milky Way. \textbf{Right-hand panel:} as above, but median [Fe/H] as a function of age. Measurements for the Milky Way's globular clusters are shown with horizontal error bars. This demonstrates that our corrected [Fe/H] values in \textsc{artemis} align the [Fe/H] vs age tracks with the Milky Way's \textit{in-situ} globular clusters \citep[those with higher {[}Fe/H{]} at a given age; see][]{myeong18,massari19}.}
   \label{fig:median_tracks}
\end{figure*}

\begin{figure}
  \centering
  \includegraphics[width=\columnwidth]{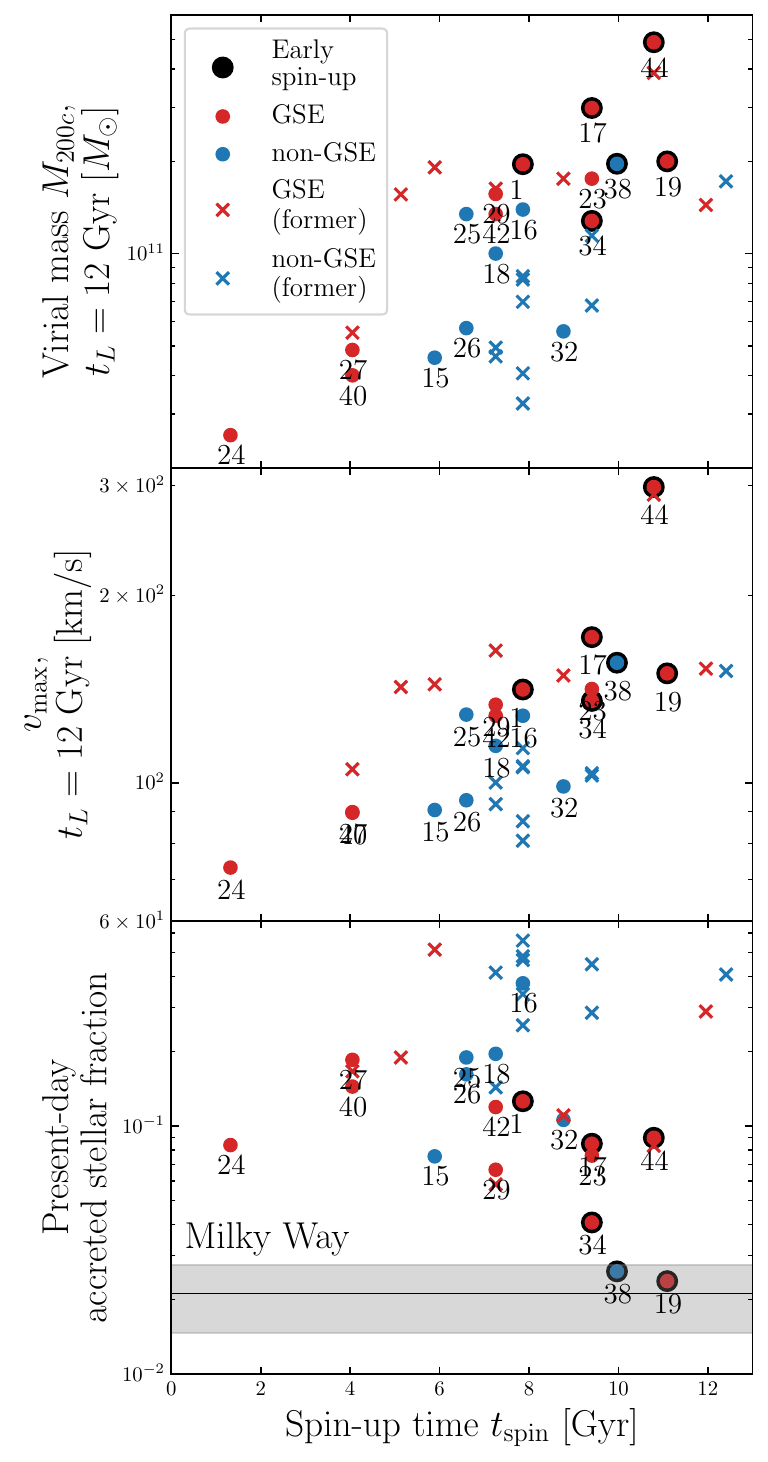}
  \caption{\textbf{Top panel:} virial mass $M_{200c}$ at a lookback time of $t_L=12$~Gyr against spin-up time $\tspin$ for all haloes with $\kappaco\geq0.5$ at the present-day (circles). Here we also show galaxies with $\kappaco\geq0.5$ at earlier snapshots only, but not at the present-day (`former discs'; crosses). Red (blue) points indicate haloes in the (non-)GSE samples, and the early spin-up galaxies selected from Fig.~\ref{fig:median_tracks} are marked with black rings. \textbf{Middle panel:} as above, but showing the maximum circular velocity $v_\mathrm{max}$ (at $t_L=12$~Gyr) against $\tspin$. In both cases there is a correlation with spin-up time, with earlier spin-ups occurring in galaxies with higher masses and circular velocities at $t_L=12$~Gyr. \textbf{Bottom panel:} fraction of accreted stars within $r=30$~kpc at the present-day against spin-up time. An estimate and its uncertainty for the Milky Way is shown with a black line and grey band. Disc galaxies with earlier spin-ups tend to have lower accreted stellar fractions, while the former discs (crosses) have much larger proportions of accreted stars.} 
   \label{fig:mass_spinup}
\end{figure}

\section{Method}
\label{section:method}

\subsection{Selection of disc galaxies}

We define a disc galaxy using the corotation parameter $\kappaco$ \citep{correa17}, the fraction of stellar kinetic energy associated with motion in the positive sense about the $z$-axis. Specifically,

\begin{equation}
    \kappaco=\frac{K_\mathrm{rot}}{K}=\frac{1}{K}\sum_i^{\substack{L_{z,i}>0, \\r<30\mathrm{kpc}}}\frac{1}{2}m_i\left(\frac{L_{z,i}}{m_iR_i}\right)^2,
\end{equation}
where $m_i$ is the mass of star particle $i$, while $L_{z,i}$ and $R_i$ are its $z$-angular momentum and projected galactocentric radius in the $x$-$y$ plane respectively. The sum is over all particles with $r<30$ kpc and positive angular momentum (i.e. circulating in the same sense as the net rotation). $K$ is the total kinetic energy of stars with $r<30$~kpc.

In \citet{dillamore22_artemis} we defined a disc galaxy using threshold $\kappaco\geq0.4$ \citep{correa17}, and used this method to identify 27 of the 45 \textsc{artemis} galaxies with discs at the present-day. In this study, we use a stricter threshold of $\kappaco\geq0.5$, which removes the borderline cases of weak discs and retains only galaxies in which more than half of the stellar kinetic energy is associated with net rotation. This results in a sample of 18 galaxies which we use in this study. In three of these $\kappaco$ decreases below 0.5 before recovering to exceed it again after $\tspin$. There are also 17 examples of galaxies where $\kappaco\geq0.5$ at some earlier snapshot but not at the present day. We include these in Fig.~\ref{fig:mass_spinup}.

Since $\kappaco$ defines the presence of a stellar disc, we also use it to define the lookback time of disc emergence or spin-up time, $\tspin$. We let $\tspin$ be the lookback time of the earliest snapshot at which $\kappaco\geq0.5$. Since $\kappaco$ takes into account all stars present within $r=30$~kpc at a particular snapshot, this should be seen as the time at which the galaxy becomes disc-dominated rather than when the disc starts forming. We therefore find that star formation in a disc-like configuration generally begins a few snapshots (or $\approx 1-2$ Gyr) before $\tspin$, as can be seen in Figure~\ref{fig:SFR_density} in the Appendix.
We also note that this figure shows that, generally, stars born before disc formation have very irregular distribution, consistent with results of \citet[][see their Fig. 12]{belokurov22} based on the FIRE-2 simulations. Thus, simulations do not show the existence of `ancient' low-metallicity discs before the actual disc spin-up at metallicities $\rm [Fe/H]\gtrsim -1.5$.

\subsection{Selection of the GSE sample}

In \citet{dillamore22_artemis} we identified \textsc{artemis} galaxies containing a feature resembling the Gaia Sausage-Enceladus (GSE) in their accreted stellar halo. This selection process is briefly described below; see \citet{dillamore22_artemis} for full details and discussion of the fitting procedure.

Particles in \textsc{artemis} are classed as \textit{in-situ} if they were born bound to the host subhalo, defined as the most massive subhalo identified by the SUBFIND algorithm \citep{Dolag_subfind} at each simulation snapshot. They are otherwise classed as accreted.

We calculate the velocity components in spherical coordinates of accreted stars in a Solar neighbourhood-like region ($5<R$~[kpc]~$<15$, $0<|z|$~[kpc]~$<9$) in each galaxy at the redshift $z=0$ snapshot. A two-component Gaussian mixture model is fitted to these velocity distributions using the expectation-maximization algorithm \textsc{GaussianMixture} from the \textsc{scikit-learn} library \citep{scikit-learn}. All stars within the above cuts are included in these fits. The two Gaussian components represent a radially anisotropic and a more isotropic distribution. The radial anisotropy is characterised by the anisotropy parameter $\beta\equiv1-(\sigma_\theta^2+\sigma_\phi^2)/(2\sigma_r^2)$ \citep{Binney_Tremaine}. For the observed GSE feature in the Milky Way, $\beta\approx0.86$ \citep{Fattahi}.

We include a halo in our GSE sample if the more anisotropic (higher $\beta$) Gaussian component has $\beta>0.8$ and a contribution to the accreted Solar neighbourhood population of greater than 40\%.  This follows a similar criteria used by \citet{Fattahi}. The GSE sample therefore consists of galaxies with significant, highly radial features in their accreted haloes. This gives us a total of 23 galaxies in the GSE sample, 10 of which also have $\kappaco\geq0.5$ at the present-day.

Our final sample of 18 disc galaxies therefore contains 10 GSE galaxies. We refer to the remaining 8 as non-GSE galaxies; these are galaxies with $\kappaco\geq0.5$ at the present-day but no GSE-like feature.

\section{Results}
\label{section:results}

\subsection{Rotational velocity, age and [Fe/H]}

We follow \citet{belokurov22} and \citet{semenov23} and investigate the relations between the median azimuthal velocity $v_\phi$, age and [Fe/H] of stars in the solar neighbourhood. We work in a coordinate system with its $z$-axis aligned with the total angular momentum of stars within $r=30$ kpc of the galactic centre. We select stars from the redshift $z=0$ snapshot from the region $5<R$~[kpc]~$<11$, $|z|<3$~kpc which are flagged as \textit{in-situ} in origin. To correct for the metallicity differences between different galaxies, we subtract the median [Fe/H] of each sample such that the corrected median [Fe/H] is zero in each galaxy. This means that the median solar neighbourhood metallicities approximately match that of the Milky Way \citep{haywood01}.

For each galaxy we divide the stars into bins of age and [Fe/H], and calculate the median $v_\phi$ and [Fe/H] in each bin. The results are shown in Fig.~\ref{fig:median_tracks}. The galaxies are divided into GSE (red) and non-GSE (blue) samples, with the medians and 16th-84th percentile ranges of each shown by the coloured lines and bands respectively. For comparison we show data for the Milky Way from APOGEE DR17 \citep[middle panel;][]{apogee,apogee_dr17,belokurov22} and the Milky Way's globular clusters \citep[right-hand panel;][]{vandenberg13}.

\citet{belokurov22} found that the transition from low to high $v_\phi$ (spin-up) generally occurs at higher metallicities in the FIRE simulations than in the Milky Way (except for the galaxy in the Romeo halo; see \citealt{mccluskey23}). The middle panel of Fig.~\ref{fig:median_tracks} shows that this is also true in the \textsc{artemis} galaxies, with a median $v_\phi=150$~km/s occuring at [Fe/H]~$\approx-0.5$ compared to -1 in the Milky Way. However, there is a difference between the GSE and non-GSE samples. The galaxies which spin up at lower metallicities and earlier times are more likely to contain a GSE analogue, whereas the median $v_\phi$ of the non-GSE galaxies spins up at later times and higher metallicities.

From the middle panel we select six galaxies which have the most Milky Way-like $v_\phi$ vs [Fe/H] tracks, henceforth `early spin-up galaxies'. These are the six galaxies with the highest median $v_\phi$ at [Fe/H]~$=-1$, five of which are in the GSE sample. The tracks of these galaxies are shown with dotted lines in each panel of Fig.~\ref{fig:median_tracks}. The left panel confirms that they do indeed spin up at earlier than average times, reaching a median $v_\phi\sim150$~km/s by $t_L\sim10$~Gyr.

\begin{figure}
  \centering
  \includegraphics[width=0.99\columnwidth]{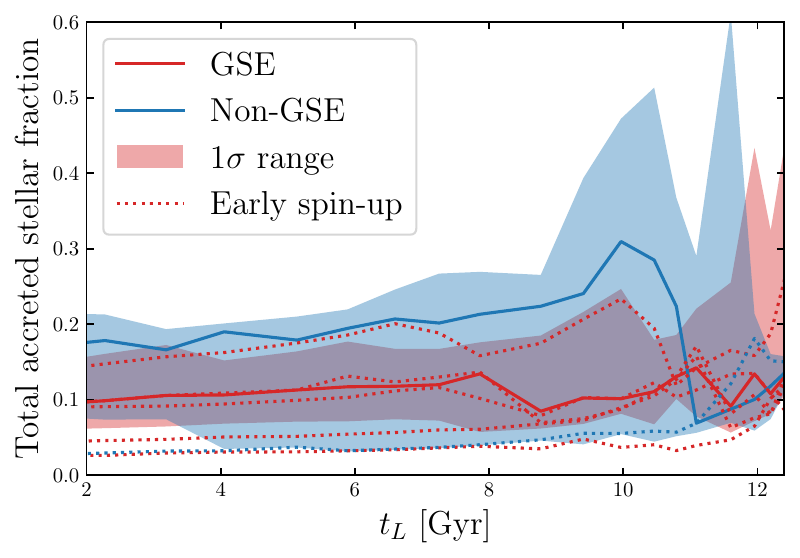}
  \caption{Accreted fraction of stars within $r=30$~kpc as a function of lookback time. In each snapshot all accreted stars present in the host galaxy at that time (within $30$~kpc) are included in the calculations. The lines and bands have the same meanings as in Fig.~\ref{fig:median_tracks}. At $t_L<11$~Gyr the non-GSE galaxies have a higher median fraction of accreted stars, with a median of about 30\% at $t_L=10$~Gyr. With one exception the early spin-up galaxies generally have low accreted fractions ($\lesssim0.1$).} 
   \label{fig:accreted_fraction}
\end{figure}

\begin{figure}
  \centering
  \includegraphics[width=0.99\columnwidth]{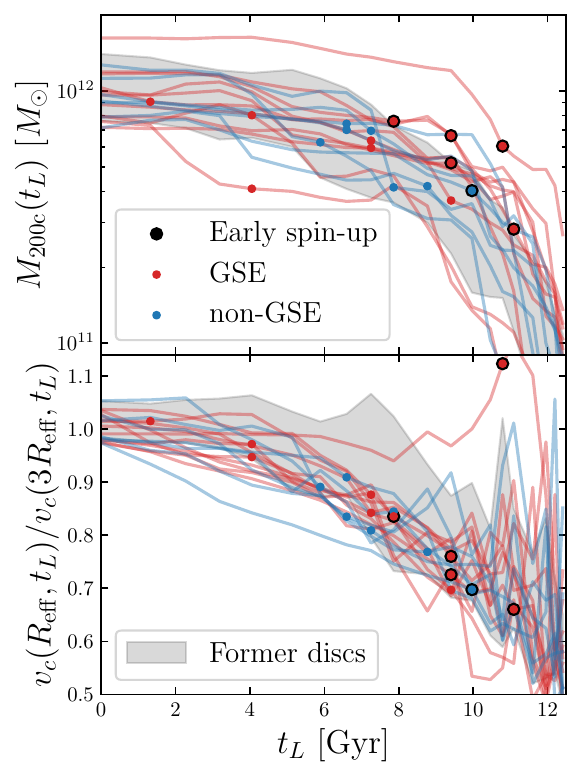}
  \caption{\textbf{Top panel:} Virial mass $M_{\rm 200c}$ vs time for the 18 present-day disc galaxies (coloured lines). The spin-up times and masses at the corresponding snapshots are shown by the coloured points, where the colours have the same meanings as in Fig.~\ref{fig:mass_spinup}. The population of former discs is also shown with the grey band, which spans between the 16th and 84th percentiles of $M_{\rm 200c}$ as a function of time. The panel shows that disc in the \textsc{artemis} galaxies form when their haloes reach mass of $M_{\rm 200c}\approx (6\pm 3)\times 10^{11}\, M_\odot$. \textbf{Bottom panel:} the ratio of the circular velocity $v_{\rm c}=[GM_{\rm tot}(<r)/r]^{0.5}$ at the stellar half-mass radius $R_\mathrm{eff}$ to that at $3R_\mathrm{eff}$, which we use as a measure of mass concentration and steepness of the potential. The panel shows that discs in the \textsc{artemis} galaxies generally form when the mass is not very centrally concentrated ($v_{\rm c}(R_{\rm eff})/v_{\rm c}(3R_{\rm eff})<1$), but the mass distribution becomes centrally concentrated after the disc forms. The former disc galaxies generally behave similarly to the present-day discs.}
   \label{fig:M200c_vcirc_ratio}
\end{figure}

\begin{figure}
  \centering
  \includegraphics[width=0.99\columnwidth]{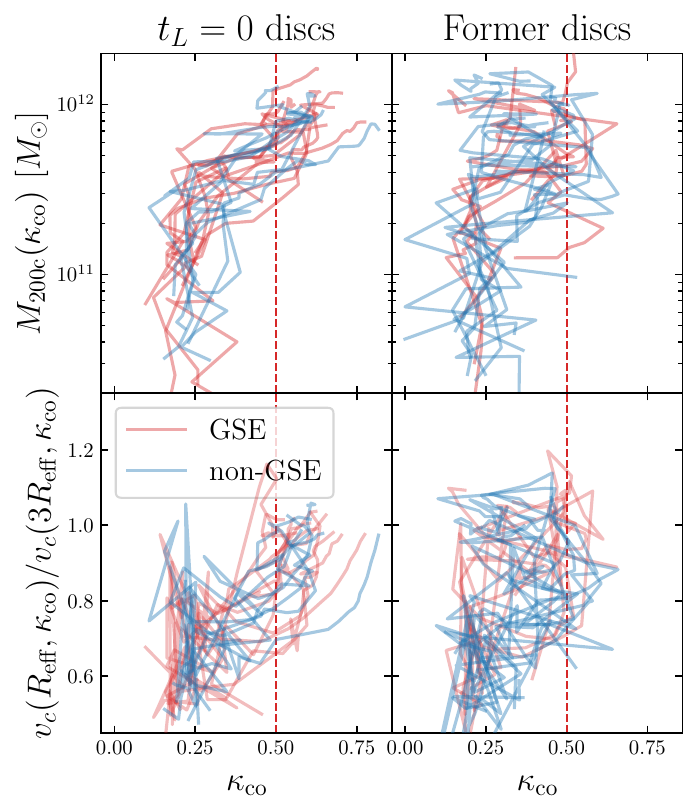}
  \caption{Like Fig.~\ref{fig:M200c_vcirc_ratio}, but $M_{\rm 200c}$ and the circular velocity ratio are plotted as functions of $\kappaco$. The left-hand column shows present-day discs, and the right column shows former discs. The spin-up threshold of $\kappaco=0.5$ is marked with a red dashed line. This confirms that spin-up occurs when the virial mass reaches the range $M_{\rm 200c}\approx (6\pm 3)\times 10^{11}$, whereas the circular velocity ratios span a wide range of values when $\kappaco$ crosses 0.5. The behaviour is generally more chaotic for the former discs, in which $\kappaco$ exceeds 0.5 before decreasing again.}
   \label{fig:M200c_vcirc_ratio_kappa_co}
\end{figure}

\begin{figure}
  \centering
  \includegraphics[width=0.95\columnwidth]{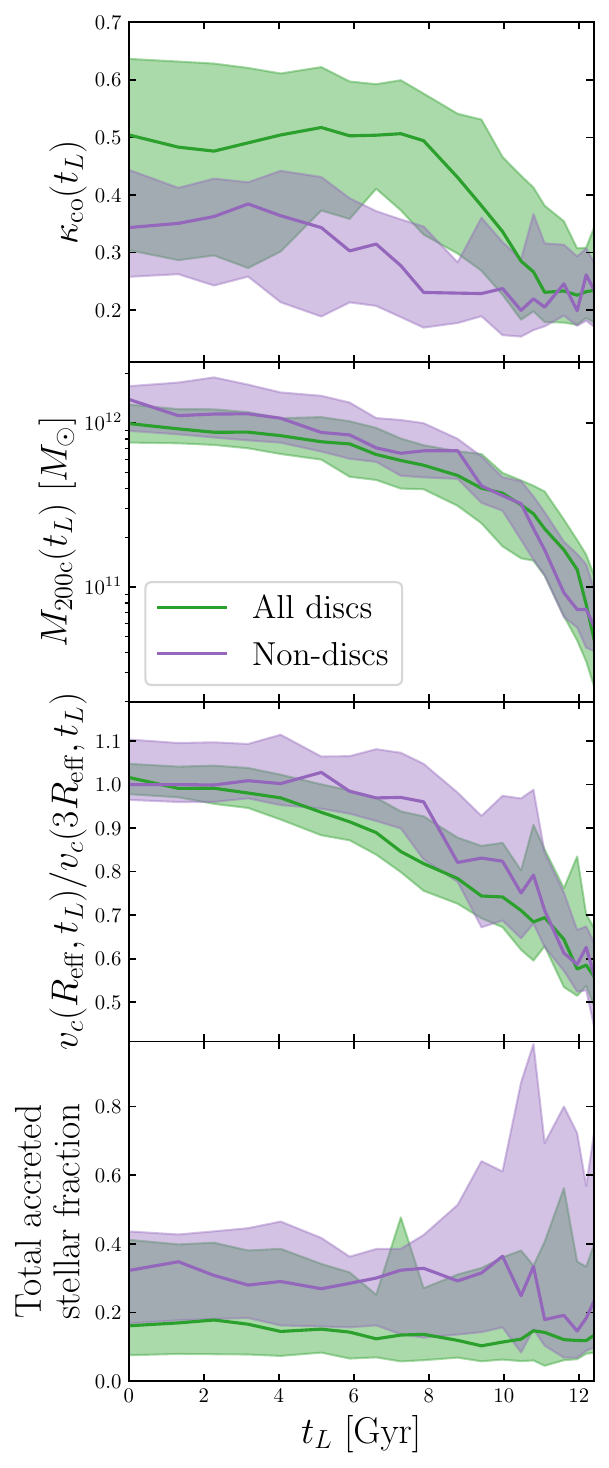}
  \caption{Various quantities against lookback time $t_L$ for galaxies with discs ($\kappaco\geq0.5$) in any snapshot (green) and those which never form discs ($\kappaco<0.5$ in every snapshot, purple). From top to bottom, the panels show the corotation parameter $\kappaco$, virial mass $M_\mathrm{200c}$, circular velocity ratio and total accreted stellar fraction (as shown in Fig.~\ref{fig:accreted_fraction}). Apart from $\kappaco$ itself, the most striking difference between the two samples is the larger fraction of accreted stars at early times in the non-disc galaxies. This suggests that the failure to form a disc may be due to merger events.}
   \label{fig:discs_vs_non_discs}
\end{figure}

\subsection{Spin-up times and evolution of haloes and their mass concentration}
\label{sec:tspin}

We show the relation between the spin-up time $\tspin$ and the galaxies' masses at early times in Figure~\ref{fig:mass_spinup}. The top and middle panels show the virial mass $M_{\rm 200c}$ and maximum circular velocity $v_\mathrm{max}$ respectively, both calculated at a lookback time of $t_L=12$~Gyr. 
The galaxies with discs ($\kappaco\geq0.5$) at the $z=0$ snapshot are shown by circles, with the early spin-up galaxies selected from Figure~\ref{fig:median_tracks} shown by circles with black edges. We also show galaxies where $\kappaco$ exceeds 0.5 at some earlier snapshot, but is less than 0.5 at the present-day (`former discs', marked with crosses). There is a moderate correlation between $\tspin$ and both the virial mass and $v_\mathrm{max}$ at $t_L=12$ Gyr, with earlier spin-ups occurring in galaxies with larger $M_{\rm 200c}(t_L=12\,\rm{Gyr})$ and $v_{\rm max}(12\,\rm{Gyr})$. In particular, the early spin-up galaxies have both larger $M_{\rm 200c}(12\,\rm{Gyr})$ and $v_{\rm max}(12\,\rm{Gyr})$ than almost all other present-day disc galaxies.

There is a large number of non-GSE galaxies with $\tspin$ between 6 and 10 Gyr and $M_{\rm 200c}(12\,\rm{Gyr})\lesssim 10^{11}M_\odot$, lower than the $M_{\rm 200c}(12\,\rm{Gyr})$ values of early spin-up galaxies' masses. This is likely related to the different accretion histories of the GSE and non-GSE galaxies. We previously showed \citep{dillamore23_artemis} that the \textsc{artemis} galaxies lacking a GSE analogue undergo more massive mergers than those with a GSE-like feature, many of which happen late. Their typical accreted satellites have stellar masses of $\sim10^{10}M_\odot$, roughly four times larger than those of GSE galaxies. This means that a greater proportion of the galaxies' final mass is assembled after $t_L=12$~Gyr. Hence to reach a final halo mass within the \textsc{artemis} selection criteria, these galaxies typically have lower masses at early times than those with a GSE analogue.

To show how accretion history relates to the spin-up time, we calculate the fraction of stars within $r=30$~kpc which are classed as accreted. This is plotted against $\tspin$ in the bottom panel of Figure~\ref{fig:mass_spinup}. For comparison, we calculate an estimate of the Milky Way's accreted fraction as follows. We take the measurement of the Milky Way's total stellar mass by \citet{mcmillan17}, $M^\star=(54.3\pm5.7)\times10^9M_\odot$, and the total stellar halo mass from \citet{deason19}, $M^\star_\mathrm{halo}=(1.4\pm0.4)\times10^9M_\odot$. We assume that a fraction $f_\mathrm{acc,halo}=0.81$ of the stellar halo is accreted \citep{naidu2020}, and calculate our estimated overall accreted fraction $f_\mathrm{acc}=f_\mathrm{halo,acc}M^\star_\mathrm{halo}/M^\star=0.021\pm 0.006$. This estimate and its uncertainty are shown in the bottom panel of Figure~\ref{fig:mass_spinup} by the black solid line and gray shaded region. Since this calculation assumes that all accreted stars are in the halo, we have checked that our results do not significantly change if we exclude accreted stars on disc-like orbits in the simulations.

For the present-day disc galaxies, the galaxies with earlier spin-ups tend to have smaller accreted stellar fractions. A large majority of the \textsc{artemis} galaxies have accreted fractions significantly larger than that of the Milky Way. The exceptions (G19, G38 and to a lesser extent G34) all have early spin-ups ($\tspin>9$~Gyr) with Milky Way-like $v_\phi$ vs [Fe/H] relations (see Fig.~\ref{fig:median_tracks}). The former disc galaxies (mostly lacking a GSE feature) tend to have larger accreted fractions, including those with early spin-ups. This suggests that mergers are responsible for destroying these discs, and is consistent with the conclusions of \citet{dillamore23_artemis} that non-GSE galaxies on average undergo more major mergers.

In Figure~\ref{fig:accreted_fraction} we show how the total accreted stellar fractions change across lookback time. At each snapshot, we calculate the fraction of all stars within $r=30$~kpc which have been accreted before that time. The medians and 16th-84th percentile ranges are plotted vs. lookback time $t_L$ for the GSE and non-GSE present-day disc galaxy samples, as well as individually for the six early spin-up galaxies. The GSE galaxies have lower median accreted fractions at all times after $t_L\approx11$~Gyr. The non-GSE galaxies have much larger contributions from accreted stars, peaking at $\approx0.3$ at $t_L=10$~Gyr. This is likely a reflection of their more active accretion histories \citep{dillamore23_artemis}. Five of the six early spin-up galaxies have small accreted fractions across all ages, at $\lesssim0.1$ at most snapshots. Finally, we note that the median accreted fractions of $\sim0.1-0.3$ are roughly consistent with the results of \citet{semenov23}.

In the top panel of Figure~\ref{fig:M200c_vcirc_ratio} we show the evolution of the virial mass $M_{\rm 200c}$ with lookback time $t_L$ for the 18 present-day disc galaxies. We mark the spin-up times and corresponding virial masses with coloured points. The 16th-84th percentile range of the former discs is also shown as a function of time. The bottom panel of Figure~\ref{fig:M200c_vcirc_ratio} shows the ratio of the circular velocity $v_{\rm c}\equiv[GM_{\rm tot}(<r)/r]^{1/2}$ at the stellar half-mass radius, $v_{\rm c}(R_\mathrm{eff})$, and  $v_{\rm c}(3R_\mathrm{eff})$, again marking the corresponding values at the disc spin-up. In \citet[][see their Figs. 6 and 7]{Hopkins23} the galaxies that form gas discs all have $v_{\rm c}(R_\mathrm{eff})/v_{\rm c}(3R_\mathrm{eff})\gtrsim 1$ at the present-day, and mass concentration (represented by the shape of the circular velocity profile) was used as a measure of the steepness of the central potential. 

Figure~\ref{fig:M200c_vcirc_ratio} shows that the virial masses at spin-up span a rather narrow range of $\sim3-9\times10^{11}M_\odot$, considerably smaller than the range of masses $M_{\rm 200c}(t_L=12\,\rm Gyr)$ in Figure~\ref{fig:mass_spinup}. Moreover, there is no apparent trend of $M_{\rm 200c}(t_{\rm spin})$ with $t_{\rm spin}$. This indicates that in the \textsc{artemis} simulations discs in $M_{\rm 200c}(z=0)\approx 10^{12}\, M_\odot$ haloes form when their host halo reaches $M_{\rm 200c}\approx (6\pm 3)\times 10^{11}\,M_\odot$. We see this more clearly in the top-left panel of Fig.~\ref{fig:M200c_vcirc_ratio_kappa_co}, where we plot $M_{\rm 200c}$ vs $\kappaco$ for each individual present-day disc galaxy. $\kappaco$ remains small ($<0.3$) while $M_{\rm 200c}$ increases from $\sim10^{10}M_\odot$ to $\sim10^{11}M_\odot$, but increases and crosses $\kappaco=0.5$ when $M_{\rm 200c}\approx(6\pm 3)\times 10^{11}\,M_\odot$.

Interestingly, this virial mass value is close to the mass threshold at which a hot halo forms and gas accretion transitions from cold-mode to hot-mode accretion \citep{Keres.etal.2005,Keres.etal.2009,Dekel.Birnboim.2006,Dekel.etal.2009}. Thus, this result is consistent with the scenario in which a hot halo is essential for disc formation by mediating the accretion of gas \citep[e.g.,][]{Dekel.etal.2020,Stern.etal.2021,Hafen.etal.2022}. Indeed, Figure 12 in \citet{Semenov.etal.2023b} shows that a hot gaseous halo forms over a significant fraction of the virial radius immediately before disc spin-up. 

At the same time, the bottom panel of Figure~\ref{fig:M200c_vcirc_ratio} shows that disc spin-up generally occurs when $v_{\rm c}(R_\mathrm{eff})/v_{\rm c}(3R_\mathrm{eff})<1$ -- i.e., when the mass distribution was not particularly centrally concentrated. This is also seen in the lower panels of Fig.~\ref{fig:M200c_vcirc_ratio_kappa_co}, where the circular velocity ratio is plotted as a function of $\kappaco$. This result is consistent with findings of \citet[][see their Figs.~7, 8 and 9]{Semenov.etal.2023b} that the mass distribution becomes more concentrated after the disc forms and that halo mass at spin-up has a narrower range than their mass concentration indicator. 

We note that all discs have $v_{\rm c}(R_\mathrm{eff})/v_{\rm c}(3R_\mathrm{eff})\gtrsim 1$ by $z=0$, in agreement with findings of \citet[][]{Hopkins23}, who found that all simulated galaxies that form a disc have $v_{\rm c}(R_\mathrm{eff})/v_{\rm c}(3R_\mathrm{eff})\gtrsim 1$ at $z=0$ (see, e.g., their Figs. 4--7). Figure~\ref{fig:M200c_vcirc_ratio}, however, shows that the mass concentration increases both before {\it and} after the disc forms and is thus not a cause of disc spin-up. 

\citet{Hopkins23} also showed that a gas disc can form in a halo with $z=0$ mass of $M_{\rm 200c}=4\times 10^{10}\, M_\odot$ (their {\tt m11a} object), which is well outside the range of  $M_{\rm 200c}(t_{\rm spin})=(6\pm 3)\times 10^{11}\, M_\odot$ found in the \textsc{artemis} simulations and the virial mass threshold expected for the hot halo formation. Nevertheless, this does not necessarily imply a discrepancy. \citet{Hopkins23} focus on the formation of \textit{gas} discs which may not necessarily lead to the formation of a stellar disc. Our definition of a disc, on the other hand, is based purely on stellar kinematics. Our findings are therefore not inconsistent with the formation of gas discs in dwarf galaxies. In fact, \citet{Hopkins23} also find that {\it at the time of formation of gaseous discs} the circular velocity is decreasing with decreasing radius at $r<3R_{\rm eff}$ and increases generally {\it after} disc formation due to the growth of central mass concentration. If stellar discs were also found to be present in lower mass galaxies, this may indicate that a hot gaseous inner halo mediating gas accretion and inducing disc formation may form in different ways. It may form when the host halo has a regular mass concentration and reaches a threshold of $M_{\rm 200c}(t_{\rm spin})=(6\pm 3)\times 10^{11}\, M_\odot$ {\it or} it may form in haloes of lower virial mass if a strong central mass concentration capable of compressing gas to a sufficiently high temperature forms via some physical process.  

These possibilities warrant further investigation. Our results, however, indicate that regular haloes with typical concentrations form discs when their central potential is not particularly steep. 

\subsection{Comparison to galaxies without discs}

So far we have focused on the \textsc{artemis} galaxies that form discs (i.e. reach $\kappaco\geq0.5$) at some snapshot. We now compare these to the general properties of those which do not reach this threshold at any time. Note that only 10 \textsc{artemis} galaxies fail to become discs at any time, so the following analysis is based on a reasonably small sample.

In Fig.~\ref{fig:discs_vs_non_discs} we show the corotation parameter $\kappaco$, virial mass $M_\mathrm{200c}$, circular velocity ratio and accreted stellar fraction vs lookback time $t_L$. As in Figs.~\ref{fig:median_tracks} and \ref{fig:accreted_fraction}, the coloured lines and bands correspond to the median and 16th-84th percentile ranges of the quantities as a function of $t_L$ for the two samples. Galaxies with $\kappaco\geq0.5$ in any snapshot are shown in green and those without are in purple.

The $\kappaco$ dependence shows the expected behaviour. The disc galaxies reach $\kappaco\sim0.5$ by $t_L\sim8$~Gyr, with some decreasing below 0.5 later (these are the `former discs'). The non-disc galaxies have significantly lower values of $\kappaco$, never exceeding 0.5 by construction. The virial mass behaviour is however similar between the two samples. This suggests that while the virial mass at early times determines the epoch of disc formation (see Figs.~\ref{fig:mass_spinup}, \ref{fig:M200c_vcirc_ratio} and \ref{fig:M200c_vcirc_ratio_kappa_co}), it does not dictate whether or not a disc forms. The same can be said of the circular velocity ratio, although the non-discs do have slightly higher values on average in most snapshots.

The most significant difference between the two samples can be seen in the accreted stellar fraction. At $t_L\approx10-12$~Gyr the median accreted fraction in non-disc galaxies increases to more than double that in those which form discs, with the 84th percentile being much higher. This period corresponds to the time at which the discs are forming (top panel and Fig.~\ref{fig:median_tracks}). This is a strong indication that the failure to form a dominant disc may be due to massive mergers at $t_L\sim10$~Gyr.

\section{Summary and conclusions}
\label{section:summary}

We have used the \textsc{artemis} simulations of galaxies in Milky Way-mass haloes to investigate the spin-up of discs. We have focused on the relations between the time of disc spin-up and the mass assembly histories of parent haloes, including the presence of a merger and associated kinematic signatures similar to the Gaia Sausage-Enceladus (GSE) merger in the Milky Way. Our results and conclusions can be summarised as follows.

\begin{enumerate}[label=\textbf{(\roman*)}]
    \item Most galaxies with discs at $z=0$ form their discs at higher metallicity than the Milky Way. \textsc{artemis} galaxies with a GSE-like feature form discs earlier and at lower metallicities than those without, and are closer to the track of the median $v_\phi-\rm [Fe/H]$ of the Milky Way (see Fig.~\ref{fig:median_tracks}). We select six early spin-up galaxies which are similar to the Milky Way (those with the highest median $v_\phi$ at [Fe/H]~$=-1$), five of which have a GSE-like feature.\\[-2mm]

    \item The distribution of stars before the disc forms is irregular without a well-defined disc or flattening (see Fig.~\ref{fig:SFR_density}). There is thus no evidence in simulations for the existence of an ancient metal-poor disc before the main disc spin-up. \\[-2mm]
    
    \item There is a correlation between spin-up time $\tspin$ and halo mass at a lookback time of $t_L=12$ Gyrs: earlier spin-ups occur in galaxies with larger $M_{\rm 200c}$ and $v_{\rm max}$ at $t_L=12$~Gyr (see top and middle panels of Fig.~\ref{fig:mass_spinup}). In particular, the six early spin-up galaxies have the largest halo masses and $v_{\rm max}$ at that epoch and experience the earliest spin-ups ($\tspin>8$~Gyr).\\[-2mm]

    \item The spin-up time also anti-correlates with the fraction of accreted stars at $z=0$. Galaxies with early spin-ups and discs tend to have smaller accreted fractions than those with later spin-ups (bottom panel of Fig.~\ref{fig:mass_spinup} and Fig.~\ref{fig:accreted_fraction}). We estimate the accreted stellar mass fraction in the Milky Way, and find that only two \textsc{artemis} galaxies have comparable fractions. Both of these have Milky Way-like spin-ups (at early times and low metallicities) and are both massive at early times. Galaxies with a GSE-like feature also tend to have lower accreted fractions, supporting the conclusions of \citet{dillamore23_artemis} that these undergo fewer major mergers than those without.\\[-2mm]

    \item We show that discs in \textsc{artemis} galaxies form when their haloes reach masses of $M_{\rm 200c}\approx (6\pm 3)\times 10^{11}\, M_\odot$ (top panel of Fig.~\ref{fig:M200c_vcirc_ratio}), similar to the mass at which galaxies are expected to form hot gaseous haloes and transition from the cold- to hot-mode accretion regime. We also show that the mass distribution becomes concentrated {\it after} the disc forms (see bottom panel of Fig.~\ref{fig:M200c_vcirc_ratio} and associated discussion in Section~\ref{sec:tspin}), suggesting that it is not the cause of disc formation.\\[-2mm]
    
    \item While the virial mass is related to the time at which the disc forms, it does not exclusively determine whether or not disc formation happens at all. The galaxies which never form discs have similar mass growth profiles to those which do, but a significantly larger proportion of their stellar populations are accreted, particularly at early times ($t_L\sim10$~Gyr). This suggests that the absence of dominant disc formation may be linked to massive mergers occurring at these times.
    \\[-4mm] 
\end{enumerate}
It is clear that the Milky Way experienced mass assembly and disc formation earlier than average for haloes of its mass. Our results show that the accretion histories of Milky Way analogues are closely linked to the formation of their discs. 

\section*{Acknowledgements}

We thank the anonymous referee for a helpful report that has improved this manuscript. We are grateful to Vadim Semenov for useful comments on the draft of this paper. 
AMD thanks the Science and Technology Facilities Council (STFC) for a PhD studentship.
This work was funded by UKRI grant 2604986. AK was supported by the National Science Foundation grants AST-1714658 and AST-1911111 and NASA ATP grant 80NSSC20K0512. For the purpose of open access, the author has applied a Creative Commons Attribution (CC BY) licence to any Author Accepted Manuscript version arising.

\section*{Data Availability}

Data from the \textsc{artemis} simulations may be shared on reasonable request to the corresponding author.



\bibliographystyle{mnras}
\bibliography{bibliography} 

\begin{thebibliography}{}
\makeatletter
\relax
\def\mn@urlcharsother{\let\do\@makeother \do\$\do\&\do\#\do\^\do\_\do\%\do\~}
\def\mn@doi{\begingroup\mn@urlcharsother \@ifnextchar [ {\mn@doi@}
  {\mn@doi@[]}}
\def\mn@doi@[#1]#2{\def\@tempa{#1}\ifx\@tempa\@empty \href
  {http://dx.doi.org/#2} {doi:#2}\else \href {http://dx.doi.org/#2} {#1}\fi
  \endgroup}
\def\mn@eprint#1#2{\mn@eprint@#1:#2::\@nil}
\def\mn@eprint@arXiv#1{\href {http://arxiv.org/abs/#1} {{\tt arXiv:#1}}}
\def\mn@eprint@dblp#1{\href {http://dblp.uni-trier.de/rec/bibtex/#1.xml}
  {dblp:#1}}
\def\mn@eprint@#1:#2:#3:#4\@nil{\def\@tempa {#1}\def\@tempb {#2}\def\@tempc
  {#3}\ifx \@tempc \@empty \let \@tempc \@tempb \let \@tempb \@tempa \fi \ifx
  \@tempb \@empty \def\@tempb {arXiv}\fi \@ifundefined
  {mn@eprint@\@tempb}{\@tempb:\@tempc}{\expandafter \expandafter \csname
  mn@eprint@\@tempb\endcsname \expandafter{\@tempc}}}

\bibitem[\protect\citeauthoryear{{Abdurro'uf} et~al.,}{{Abdurro'uf}
  et~al.}{2022}]{apogee_dr17}
{Abdurro'uf} et~al., 2022, \mn@doi [\apjs] {10.3847/1538-4365/ac4414}, \href
  {https://ui.adsabs.harvard.edu/abs/2022ApJS..259...35A} {259, 35}

\bibitem[\protect\citeauthoryear{{Belokurov} \& {Kravtsov}}{{Belokurov} \&
  {Kravtsov}}{2022}]{belokurov22}
{Belokurov} V.,  {Kravtsov} A.,  2022, \mn@doi [\mnras]
  {10.1093/mnras/stac1267}, \href
  {https://ui.adsabs.harvard.edu/abs/2022MNRAS.514..689B} {514, 689}

\bibitem[\protect\citeauthoryear{{Belokurov}, {Erkal}, {Evans}, {Koposov}  \&
  {Deason}}{{Belokurov} et~al.}{2018}]{belokurov18}
{Belokurov} V.,  {Erkal} D.,  {Evans} N.~W.,  {Koposov} S.~E.,   {Deason}
  A.~J.,  2018, \mn@doi [\mnras] {10.1093/mnras/sty982}, \href
  {https://ui.adsabs.harvard.edu/abs/2018MNRAS.478..611B} {478, 611}

\bibitem[\protect\citeauthoryear{{Belokurov}, {Sanders}, {Fattahi}, {Smith},
  {Deason}, {Evans}  \& {Grand}}{{Belokurov} et~al.}{2020}]{belokurov2020}
{Belokurov} V.,  {Sanders} J.~L.,  {Fattahi} A.,  {Smith} M.~C.,  {Deason}
  A.~J.,  {Evans} N.~W.,   {Grand} R. J.~J.,  2020, \mn@doi [\mnras]
  {10.1093/mnras/staa876}, \href
  {https://ui.adsabs.harvard.edu/abs/2020MNRAS.494.3880B} {494, 3880}

\bibitem[\protect\citeauthoryear{{Binney} \& {Tremaine}}{{Binney} \&
  {Tremaine}}{2008}]{Binney_Tremaine}
{Binney} J.,  {Tremaine} S.,  2008, {Galactic Dynamics: Second Edition}

\bibitem[\protect\citeauthoryear{{Conroy} et~al.,}{{Conroy}
  et~al.}{2022}]{Conroy22}
{Conroy} C.,  et~al., 2022, \mn@doi [arXiv e-prints]
  {10.48550/arXiv.2204.02989}, \href
  {https://ui.adsabs.harvard.edu/abs/2022arXiv220402989C} {p. arXiv:2204.02989}

\bibitem[\protect\citeauthoryear{{Correa}, {Schaye}, {Clauwens}, {Bower},
  {Crain}, {Schaller}, {Theuns}  \& {Thob}}{{Correa} et~al.}{2017}]{correa17}
{Correa} C.~A.,  {Schaye} J.,  {Clauwens} B.,  {Bower} R.~G.,  {Crain} R.~A.,
  {Schaller} M.,  {Theuns} T.,   {Thob} A. C.~R.,  2017, \mn@doi [\mnras]
  {10.1093/mnrasl/slx133}, \href
  {https://ui.adsabs.harvard.edu/abs/2017MNRAS.472L..45C} {472, L45}

\bibitem[\protect\citeauthoryear{{Crain} et~al.,}{{Crain}
  et~al.}{2015}]{crain15}
{Crain} R.~A.,  et~al., 2015, \mn@doi [\mnras] {10.1093/mnras/stv725}, \href
  {https://ui.adsabs.harvard.edu/abs/2015MNRAS.450.1937C} {450, 1937}

\bibitem[\protect\citeauthoryear{{De Silva} et~al.,}{{De Silva}
  et~al.}{2015}]{Galah}
{De Silva} G.~M.,  et~al., 2015, \mn@doi [\mnras] {10.1093/mnras/stv327}, \href
  {https://ui.adsabs.harvard.edu/abs/2015MNRAS.449.2604D} {449, 2604}

\bibitem[\protect\citeauthoryear{{Deason}, {Belokurov}  \& {Sanders}}{{Deason}
  et~al.}{2019}]{deason19}
{Deason} A.~J.,  {Belokurov} V.,   {Sanders} J.~L.,  2019, \mn@doi [\mnras]
  {10.1093/mnras/stz2793}, \href
  {https://ui.adsabs.harvard.edu/abs/2019MNRAS.490.3426D} {490, 3426}

\bibitem[\protect\citeauthoryear{{Dekel} \& {Birnboim}}{{Dekel} \&
  {Birnboim}}{2006}]{Dekel.Birnboim.2006}
{Dekel} A.,  {Birnboim} Y.,  2006, \mn@doi [\mnras]
  {10.1111/j.1365-2966.2006.10145.x}, \href
  {https://ui.adsabs.harvard.edu/abs/2006MNRAS.368....2D} {368, 2}

\bibitem[\protect\citeauthoryear{{Dekel} et~al.,}{{Dekel}
  et~al.}{2009}]{Dekel.etal.2009}
{Dekel} A.,  et~al., 2009, \mn@doi [\nat] {10.1038/nature07648}, \href
  {https://ui.adsabs.harvard.edu/abs/2009Natur.457..451D} {457, 451}

\bibitem[\protect\citeauthoryear{{Dekel}, {Ginzburg}, {Jiang}, {Freundlich},
  {Lapiner}, {Ceverino}  \& {Primack}}{{Dekel} et~al.}{2020}]{Dekel.etal.2020}
{Dekel} A.,  {Ginzburg} O.,  {Jiang} F.,  {Freundlich} J.,  {Lapiner} S.,
  {Ceverino} D.,   {Primack} J.,  2020, \mn@doi [\mnras]
  {10.1093/mnras/staa470}, \href
  {https://ui.adsabs.harvard.edu/abs/2020MNRAS.493.4126D} {493, 4126}

\bibitem[\protect\citeauthoryear{{Dillamore}, {Belokurov}, {Font}  \&
  {McCarthy}}{{Dillamore} et~al.}{2022}]{dillamore22_artemis}
{Dillamore} A.~M.,  {Belokurov} V.,  {Font} A.~S.,   {McCarthy} I.~G.,  2022,
  \mn@doi [\mnras] {10.1093/mnras/stac1038}, \href
  {https://ui.adsabs.harvard.edu/abs/2022MNRAS.513.1867D} {513, 1867}

\bibitem[\protect\citeauthoryear{{Dillamore}, {Belokurov}, {Evans}  \&
  {Font}}{{Dillamore} et~al.}{2023}]{dillamore23_artemis}
{Dillamore} A.~M.,  {Belokurov} V.,  {Evans} N.~W.,   {Font} A.~S.,  2023,
  \mn@doi [\mnras] {10.1093/mnrasl/slac158}, \href
  {https://ui.adsabs.harvard.edu/abs/2023MNRAS.519L..87D} {519, L87}

\bibitem[\protect\citeauthoryear{{Dolag}, {Borgani}, {Murante}  \&
  {Springel}}{{Dolag} et~al.}{2009}]{Dolag_subfind}
{Dolag} K.,  {Borgani} S.,  {Murante} G.,   {Springel} V.,  2009, \mn@doi
  [\mnras] {10.1111/j.1365-2966.2009.15034.x}, \href
  {https://ui.adsabs.harvard.edu/abs/2009MNRAS.399..497D} {399, 497}

\bibitem[\protect\citeauthoryear{{Fattahi} et~al.,}{{Fattahi}
  et~al.}{2019}]{Fattahi}
{Fattahi} A.,  et~al., 2019, \mn@doi [\mnras] {10.1093/mnras/stz159}, \href
  {https://ui.adsabs.harvard.edu/abs/2019MNRAS.484.4471F} {484, 4471}

\bibitem[\protect\citeauthoryear{{Ferreira} et~al.,}{{Ferreira}
  et~al.}{2022}]{Ferreira.etal.2022}
{Ferreira} L.,  et~al., 2022, \mn@doi [\apjl] {10.3847/2041-8213/ac947c}, \href
  {https://ui.adsabs.harvard.edu/abs/2022ApJ...938L...2F} {938, L2}

\bibitem[\protect\citeauthoryear{{Ferreira} et~al.,}{{Ferreira}
  et~al.}{2023}]{Ferreira.etal.2023}
{Ferreira} L.,  et~al., 2023, \mn@doi [\apj] {10.3847/1538-4357/acec76}, \href
  {https://ui.adsabs.harvard.edu/abs/2023ApJ...955...94F} {955, 94}

\bibitem[\protect\citeauthoryear{{Font} et~al.,}{{Font} et~al.}{2020}]{artemis}
{Font} A.~S.,  et~al., 2020, \mn@doi [\mnras] {10.1093/mnras/staa2463}, \href
  {https://ui.adsabs.harvard.edu/abs/2020MNRAS.498.1765F} {498, 1765}

\bibitem[\protect\citeauthoryear{{Gaia Collaboration} et~al.,}{{Gaia
  Collaboration} et~al.}{2016}]{Gaia}
{Gaia Collaboration} et~al., 2016, \mn@doi [\aap]
  {10.1051/0004-6361/201629272}, \href
  {https://ui.adsabs.harvard.edu/abs/2016A&A...595A...1G} {595, A1}

\bibitem[\protect\citeauthoryear{{Grand} et~al.,}{{Grand}
  et~al.}{2017}]{grand2017}
{Grand} R. J.~J.,  et~al., 2017, \mn@doi [\mnras] {10.1093/mnras/stx071}, \href
  {https://ui.adsabs.harvard.edu/abs/2017MNRAS.467..179G} {467, 179}

\bibitem[\protect\citeauthoryear{{Grand} et~al.,}{{Grand}
  et~al.}{2020}]{grand2020}
{Grand} R. J.~J.,  et~al., 2020, \mn@doi [\mnras] {10.1093/mnras/staa2057},
  \href {https://ui.adsabs.harvard.edu/abs/2020MNRAS.497.1603G} {497, 1603}

\bibitem[\protect\citeauthoryear{{Hafen} et~al.,}{{Hafen}
  et~al.}{2022}]{Hafen.etal.2022}
{Hafen} Z.,  et~al., 2022, \mn@doi [\mnras] {10.1093/mnras/stac1603}, \href
  {https://ui.adsabs.harvard.edu/abs/2022MNRAS.514.5056H} {514, 5056}

\bibitem[\protect\citeauthoryear{{Haywood}}{{Haywood}}{2001}]{haywood01}
{Haywood} M.,  2001, \mn@doi [\mnras] {10.1046/j.1365-8711.2001.04510.x}, \href
  {https://ui.adsabs.harvard.edu/abs/2001MNRAS.325.1365H} {325, 1365}

\bibitem[\protect\citeauthoryear{{Helmi}, {Babusiaux}, {Koppelman}, {Massari},
  {Veljanoski}  \& {Brown}}{{Helmi} et~al.}{2018}]{helmi18}
{Helmi} A.,  {Babusiaux} C.,  {Koppelman} H.~H.,  {Massari} D.,  {Veljanoski}
  J.,   {Brown} A. G.~A.,  2018, \mn@doi [\nat] {10.1038/s41586-018-0625-x},
  \href {https://ui.adsabs.harvard.edu/abs/2018Natur.563...85H} {563, 85}

\bibitem[\protect\citeauthoryear{{Hopkins} et~al.,}{{Hopkins}
  et~al.}{2023}]{Hopkins23}
{Hopkins} P.~F.,  et~al., 2023, \mn@doi [\mnras] {10.1093/mnras/stad1902},
  \href {https://ui.adsabs.harvard.edu/abs/2023MNRAS.525.2241H} {525, 2241}

\bibitem[\protect\citeauthoryear{{Jacobs} et~al.,}{{Jacobs}
  et~al.}{2023}]{Jacobs.etal.2023}
{Jacobs} C.,  et~al., 2023, \mn@doi [\apjl] {10.3847/2041-8213/accd6d}, \href
  {https://ui.adsabs.harvard.edu/abs/2023ApJ...948L..13J} {948, L13}

\bibitem[\protect\citeauthoryear{{Kassin} et~al.,}{{Kassin}
  et~al.}{2012}]{Kassin12}
{Kassin} S.~A.,  et~al., 2012, \mn@doi [\apj] {10.1088/0004-637X/758/2/106},
  \href {https://ui.adsabs.harvard.edu/abs/2012ApJ...758..106K} {758, 106}

\bibitem[\protect\citeauthoryear{{Kere{\v{s}}}, {Katz}, {Weinberg}  \&
  {Dav{\'e}}}{{Kere{\v{s}}} et~al.}{2005}]{Keres.etal.2005}
{Kere{\v{s}}} D.,  {Katz} N.,  {Weinberg} D.~H.,   {Dav{\'e}} R.,  2005,
  \mn@doi [\mnras] {10.1111/j.1365-2966.2005.09451.x}, \href
  {https://ui.adsabs.harvard.edu/abs/2005MNRAS.363....2K} {363, 2}

\bibitem[\protect\citeauthoryear{{Kere{\v{s}}}, {Katz}, {Fardal}, {Dav{\'e}}
  \& {Weinberg}}{{Kere{\v{s}}} et~al.}{2009}]{Keres.etal.2009}
{Kere{\v{s}}} D.,  {Katz} N.,  {Fardal} M.,  {Dav{\'e}} R.,   {Weinberg} D.~H.,
   2009, \mn@doi [\mnras] {10.1111/j.1365-2966.2009.14541.x}, \href
  {https://ui.adsabs.harvard.edu/abs/2009MNRAS.395..160K} {395, 160}

\bibitem[\protect\citeauthoryear{{Majewski} et~al.,}{{Majewski}
  et~al.}{2017}]{apogee}
{Majewski} S.~R.,  et~al., 2017, \mn@doi [\aj] {10.3847/1538-3881/aa784d},
  \href {https://ui.adsabs.harvard.edu/abs/2017AJ....154...94M} {154, 94}

\bibitem[\protect\citeauthoryear{{Massari}, {Koppelman}  \& {Helmi}}{{Massari}
  et~al.}{2019}]{massari19}
{Massari} D.,  {Koppelman} H.~H.,   {Helmi} A.,  2019, \mn@doi [\aap]
  {10.1051/0004-6361/201936135}, \href
  {https://ui.adsabs.harvard.edu/abs/2019A&A...630L...4M} {630, L4}

\bibitem[\protect\citeauthoryear{{McCluskey}, {Wetzel}, {Loebman}, {Moreno}  \&
  {Faucher-Giguere}}{{McCluskey} et~al.}{2023}]{mccluskey23}
{McCluskey} F.,  {Wetzel} A.,  {Loebman} S.~R.,  {Moreno} J.,
  {Faucher-Giguere} C.-A.,  2023, \mn@doi [arXiv e-prints]
  {10.48550/arXiv.2303.14210}, \href
  {https://ui.adsabs.harvard.edu/abs/2023arXiv230314210M} {p. arXiv:2303.14210}

\bibitem[\protect\citeauthoryear{{McMillan}}{{McMillan}}{2017}]{mcmillan17}
{McMillan} P.~J.,  2017, \mn@doi [\mnras] {10.1093/mnras/stw2759}, \href
  {https://ui.adsabs.harvard.edu/abs/2017MNRAS.465...76M} {465, 76}

\bibitem[\protect\citeauthoryear{{Myeong}, {Evans}, {Belokurov}, {Sanders}  \&
  {Koposov}}{{Myeong} et~al.}{2018}]{myeong18}
{Myeong} G.~C.,  {Evans} N.~W.,  {Belokurov} V.,  {Sanders} J.~L.,   {Koposov}
  S.~E.,  2018, \mn@doi [\apjl] {10.3847/2041-8213/aad7f7}, \href
  {https://ui.adsabs.harvard.edu/abs/2018ApJ...863L..28M} {863, L28}

\bibitem[\protect\citeauthoryear{{Naidu}, {Conroy}, {Bonaca}, {Johnson},
  {Ting}, {Caldwell}, {Zaritsky}  \& {Cargile}}{{Naidu}
  et~al.}{2020}]{naidu2020}
{Naidu} R.~P.,  {Conroy} C.,  {Bonaca} A.,  {Johnson} B.~D.,  {Ting} Y.-S.,
  {Caldwell} N.,  {Zaritsky} D.,   {Cargile} P.~A.,  2020, \mn@doi [\apj]
  {10.3847/1538-4357/abaef4}, \href
  {https://ui.adsabs.harvard.edu/abs/2020ApJ...901...48N} {901, 48}

\bibitem[\protect\citeauthoryear{{Nelson} et~al.,}{{Nelson}
  et~al.}{2023}]{Nelson23}
{Nelson} E.~J.,  et~al., 2023, \mn@doi [\apjl] {10.3847/2041-8213/acc1e1},
  \href {https://ui.adsabs.harvard.edu/abs/2023ApJ...948L..18N} {948, L18}

\bibitem[\protect\citeauthoryear{{Orkney} et~al.,}{{Orkney}
  et~al.}{2023}]{orkney2023}
{Orkney} M. D.~A.,  et~al., 2023, \mn@doi [\mnras] {10.1093/mnras/stad2361},
  \href {https://ui.adsabs.harvard.edu/abs/2023MNRAS.525..683O} {525, 683}

\bibitem[\protect\citeauthoryear{Pedregosa et~al.,}{Pedregosa
  et~al.}{2011}]{scikit-learn}
Pedregosa F.,  et~al., 2011, Journal of Machine Learning Research, 12, 2825

\bibitem[\protect\citeauthoryear{{Rix} et~al.,}{{Rix} et~al.}{2022}]{Rix22}
{Rix} H.-W.,  et~al., 2022, \mn@doi [\apj] {10.3847/1538-4357/ac9e01}, \href
  {https://ui.adsabs.harvard.edu/abs/2022ApJ...941...45R} {941, 45}

\bibitem[\protect\citeauthoryear{{Robertson} et~al.,}{{Robertson}
  et~al.}{2023}]{Robertson.etal.2023}
{Robertson} B.~E.,  et~al., 2023, \mn@doi [\apjl] {10.3847/2041-8213/aca086},
  \href {https://ui.adsabs.harvard.edu/abs/2023ApJ...942L..42R} {942, L42}

\bibitem[\protect\citeauthoryear{{Schaye} et~al.,}{{Schaye}
  et~al.}{2015}]{schaye15}
{Schaye} J.,  et~al., 2015, \mn@doi [\mnras] {10.1093/mnras/stu2058}, \href
  {https://ui.adsabs.harvard.edu/abs/2015MNRAS.446..521S} {446, 521}

\bibitem[\protect\citeauthoryear{{Semenov}, {Conroy}, {Chandra}, {Hernquist}
  \& {Nelson}}{{Semenov} et~al.}{2023a}]{semenov23}
{Semenov} V.~A.,  {Conroy} C.,  {Chandra} V.,  {Hernquist} L.,   {Nelson} D.,
  2023a, \mn@doi [arXiv e-prints] {10.48550/arXiv.2306.09398}, \href
  {https://ui.adsabs.harvard.edu/abs/2023arXiv230609398S} {p. arXiv:2306.09398}

\bibitem[\protect\citeauthoryear{{Semenov}, {Conroy}, {Chandra}, {Hernquist}
  \& {Nelson}}{{Semenov} et~al.}{2023b}]{Semenov.etal.2023b}
{Semenov} V.~A.,  {Conroy} C.,  {Chandra} V.,  {Hernquist} L.,   {Nelson} D.,
  2023b, \mn@doi [arXiv e-prints] {10.48550/arXiv.2306.13125}, \href
  {https://ui.adsabs.harvard.edu/abs/2023arXiv230613125S} {p. arXiv:2306.13125}

\bibitem[\protect\citeauthoryear{{Shen}, {Mo}, {White}, {Blanton}, {Kauffmann},
  {Voges}, {Brinkmann}  \& {Csabai}}{{Shen} et~al.}{2003}]{shen03}
{Shen} S.,  {Mo} H.~J.,  {White} S. D.~M.,  {Blanton} M.~R.,  {Kauffmann} G.,
  {Voges} W.,  {Brinkmann} J.,   {Csabai} I.,  2003, \mn@doi [\mnras]
  {10.1046/j.1365-8711.2003.06740.x}, \href
  {https://ui.adsabs.harvard.edu/abs/2003MNRAS.343..978S} {343, 978}

\bibitem[\protect\citeauthoryear{{Springel}}{{Springel}}{2005}]{springel05}
{Springel} V.,  2005, \mn@doi [\mnras] {10.1111/j.1365-2966.2005.09655.x},
  \href {https://ui.adsabs.harvard.edu/abs/2005MNRAS.364.1105S} {364, 1105}

\bibitem[\protect\citeauthoryear{{Stern} et~al.,}{{Stern}
  et~al.}{2021}]{Stern.etal.2021}
{Stern} J.,  et~al., 2021, \mn@doi [\apj] {10.3847/1538-4357/abd776}, \href
  {https://ui.adsabs.harvard.edu/abs/2021ApJ...911...88S} {911, 88}

\bibitem[\protect\citeauthoryear{{VandenBerg}, {Brogaard}, {Leaman}  \&
  {Casagrande}}{{VandenBerg} et~al.}{2013}]{vandenberg13}
{VandenBerg} D.~A.,  {Brogaard} K.,  {Leaman} R.,   {Casagrande} L.,  2013,
  \mn@doi [\apj] {10.1088/0004-637X/775/2/134}, \href
  {https://ui.adsabs.harvard.edu/abs/2013ApJ...775..134V} {775, 134}

\bibitem[\protect\citeauthoryear{{Wisnioski} et~al.,}{{Wisnioski}
  et~al.}{2015}]{Wisnioski15}
{Wisnioski} E.,  et~al., 2015, \mn@doi [\apj] {10.1088/0004-637X/799/2/209},
  \href {https://ui.adsabs.harvard.edu/abs/2015ApJ...799..209W} {799, 209}

\bibitem[\protect\citeauthoryear{{Wisnioski} et~al.,}{{Wisnioski}
  et~al.}{2019}]{Wisnioski19}
{Wisnioski} E.,  et~al., 2019, \mn@doi [\apj] {10.3847/1538-4357/ab4db8}, \href
  {https://ui.adsabs.harvard.edu/abs/2019ApJ...886..124W} {886, 124}

\makeatother
\end{thebibliography}




\appendix

\section{Star formation rate density}
\label{section:SFR}

\begin{figure*}
  \centering
  \includegraphics[width=\textwidth]{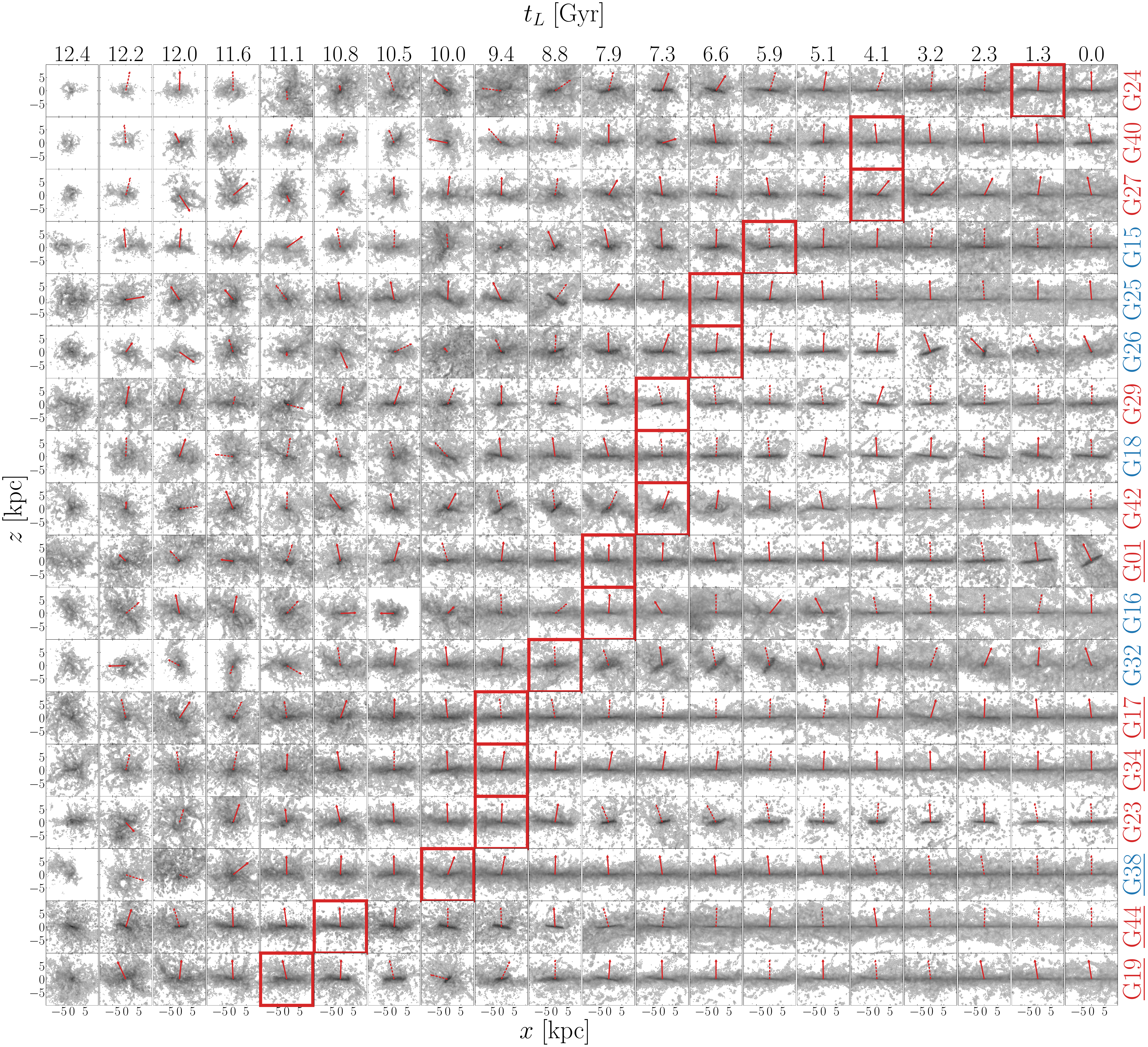}
  \caption{Star formation rate density in the inner 10~kpc of all 18 galaxies in our sample (rows) across many snapshots (columns). The z-axis in each panel is parallel to the instantaneous angular momentum of gas particles within a radius of $r=10$~kpc. For each galaxy the snapshot at the spin-up time $\tspin$ is marked with a red border, with $\tspin$ becoming earlier from top to bottom. The red arrows mark the direction of the $z$-axis in the previous snapshot, which indicates how much this direction changes between consecutive snapshots. As in the other figures the red (blue) colours of the galaxy IDs correspond to (non-)GSE galaxies, and the early spin-up galaxies highlighted in Figs.~\ref{fig:median_tracks} and \ref{fig:mass_spinup} are underlined.}
   \label{fig:SFR_density}
\end{figure*}

In Fig.~\ref{fig:SFR_density} we show the projected star formation rate density in each of the 18 present-day disc galaxies, from $t_L=12.4$ to 0~Gyr. Each row corresponds to a different galaxy, labelled on the right-hand side. The $z$-axis corresponds to the angular momentum of the gas within $r=10$~kpc in each snapshot, with the $z$ unit vector from the previous snapshot marked with a red arrow. The galaxies are ordered by $\tspin$, with the spin-up (marked with a red border) becoming earlier moving down the rows. The star formation is initially clumpy and irregular, with the angular momentum direction of the gas changing dramatically between subsequent snapshots. The stellar discs (parallel to the $x$-axis) can then be seen forming in the few snapshots prior to the spin-up time, after which star formation usually takes place only close to this plane. The relation between early-time mass and spin-up time shown in Fig.~\ref{fig:mass_spinup} can also be discerned, with rates of star formation in the first few snapshots being visibly lower in the galaxies with the latest spin-ups (top few rows).

Note that the presence of a disc (i.e. $\kappaco>0.5$) does not necessarily imply that star formation is occurring in a disc-like configuration. For example, the star formation in G26 and G32 at $t_L=0$ is more dispersed. Conversely, star formation in a disc does not necessarily imply that $\kappaco>0.5$ (e.g. G24 and G27 before spin-up). In both scenarios this may be a reflection of the fact that $\kappaco$ describes the overall state of all stars, whereas Fig.~\ref{fig:SFR_density} only shows where star formation is actively ongoing. A disc may therefore be present but not star forming (e.g. G26 and G32), or actively forming stars but not yet dominant (e.g. G24 and G27). We also find in a few cases (including G24 and G27) that $\kappaco$ approaches close to 0.5 a few Gyr before it exceeds that value. Hence their late spin-up times are partly due to their discs not becoming dominant enough until some time after their initial formation.


\bsp	
\label{lastpage}
\end{document}